\documentclass[10pt,aps,pre,twocolumn,showpacs,superscriptaddress]{revtex4-1}
\usepackage{hyperref}
\usepackage{graphicx}
\usepackage{amsmath,amssymb}
\usepackage{ulem}
\usepackage{xcolor}

\newcommand{\ins}[1]{\textcolor[rgb]{0,0,0}{#1}}
\begin{document}

\title{Synchronously controlled optical modes in the transmittance and reflectance spectra of multilayer photonic structure with dual-frequency nematic liquid crystal}

\author{Vladimir A. Gunyakov}
\email{gun@iph.krasn.ru}
\affiliation{Kirensky Institute of Physics, Federal Research Center -- Krasnoyarsk Scientific Center, Siberian Branch, Russian Academy of Sciences, Krasnoyarsk 660036, Russia}
\author{Vitaly S. Sutormin}
\affiliation{Kirensky Institute of Physics, Federal Research Center -- Krasnoyarsk Scientific Center, Siberian Branch, Russian Academy of Sciences, Krasnoyarsk 660036, Russia}
\affiliation{Siberian Federal University, Krasnoyarsk 660041, Russia}
\author{Sergey~A.~Myslivets}
\affiliation{Kirensky Institute of Physics, Federal Research Center -- Krasnoyarsk Scientific Center, Siberian Branch, Russian Academy of Sciences, Krasnoyarsk 660036, Russia}
\affiliation{Siberian Federal University, Krasnoyarsk 660041, Russia}
\author{Vasily F. Shabanov}
\affiliation{Kirensky Institute of Physics, Federal Research Center -- Krasnoyarsk Scientific Center, Siberian Branch, Russian Academy of Sciences, Krasnoyarsk 660036, Russia}
\author{Victor Ya. Zyryanov}
\affiliation{Kirensky Institute of Physics, Federal Research Center -- Krasnoyarsk Scientific Center, Siberian Branch, Russian Academy of Sciences, Krasnoyarsk 660036, Russia}

\date{\today}

\begin{abstract}
A method for the electrically controlled synchronous mode tuning in the transmittance and reflectance spectra of a photonic structure consisting of an asymmetric dielectric Fabry––P\'{e}rot microcavity and an ultrathin metallic film  has been proposed. The excitation of a broadband Tamm plasmon-polariton at the metal/Bragg mirror interface is accompanied by the two phenomena opposing one another. Due to the strong absorption induced by the metal, a sufficient rejection level in reflection of the off-resonant radiation has been obtained in a wide spectral range, which almost coincides with the photonic band gap, while at the cavity mode frequencies the structure becomes transparent for the reflected radiation. This leads to the appearance of a series of narrow resonance peaks in the reflectance spectra that coincide in frequency with the transmittance peaks. Based on the structural transformations in the dual-frequency nematic liquid crystal used as a photonic structure defect, the tuning of the modes corresponding to the extraordinary waves to both the short- and long-wavelength spectral ranges has been implemented.
\end{abstract}


\maketitle

\section{Introduction}

The Fabry–P\'{e}rot cavity-type multilayer photonic structures (PSs) based on the distributed Bragg mirrors with a defect layer between them, which are characterized by the appearance of transmission peaks in the band gaps, are the optical materials promising for creating functional elements of nanophotonic and optoelectronic devices \cite{1,2}. These peaks correspond to the spatial distributions of light fields called localized (defect) modes. In general, all-dielectric devices of this type cannot operate simultaneously in the transmission and reflection regime, since these regimes complement each other: at the mode frequencies, the transmittance peaks will correspond to dips against a wide reflection band coinciding with the photonic band gap (PBG). The use of ultrathin metallic films with a thickness much less than the light wavelength as additional elements of multilayer structures allows one not only to optimize their reflectivity \cite{3,4,5} but also to create fundamentally new structures capable of operating in the selective transmission and reflection modes simultaneously. In particular, based on the asymmetric dielectric Fabry–P\'{e}rot cavity containing an ultrathin metallic film with the refractive index $n_{m}=n-ik$, the real and imaginary parts of which should satisfy the condition $n\approx k$, a filter operating in both regimes was developed \cite{6}. 
The theoretical and experimental studies showed that the reflectance and transmittance are maximum at a central wavelength of $\lambda_{0}=700$~nm. At the other wavelengths, the metal, being an effective absorber, ensured a high rejection level in reflection. The possibility of the angular tuning of the properties of a new reflection-and-transmission filter was theoretically investigated in \cite{7}. It was demonstrated that, as the angle of incidence of the probe radiation increases, the reflection and transmission peaks synchronously shift toward shorter wavelengths for both the TM and TE polarizations. However, for the mode frequency tuning, the field control techniques are more suitable in nanophotonic and optoelectronic devices. In practice, the photonic structures with liquid-crystal (LC) components are highly promising. The properties of LCs, including their wide spectral transparency range, large birefringence, and high sensitivity to external factors (temperature, magnetic, and electric  fields) open up wide opportunities for efficient control of the spectral and optical characteristics of PSs \ins{\cite{Kitz,Oz1,Oz2,8,9,10}}. Among LCs, of particular interest are the dual-frequency cholesteric and nematic mixtures, which conventionally exhibit the positive dielectric anisotropy at low frequencies and the negative dielectric anisotropy above a certain frequency of an applied electric field. Therefore, the alignment of molecules and, consequently, the optical properties of the LC layers can be significantly changed via switching the frequency of an applied voltage \cite{11,12}, which makes it possible to use LCs as a controllable element in the reflection-and-transmission PSs.

At the same time, features of the field-induced structural transformations in the LC layer impose certain restrictions on its thickness. The layers with a thickness of several microns are considered to be optimal \cite{13}. It should be noted that the approach developed in \cite{6,7} was implemented primarily for the single-mode PSs, in which the optical thickness of a structural defect is a half-wave layer. In this case, there are single transmission and reflection peaks within the band gap, the frequency of which coincides with the Bragg frequency of a periodic structure of mirrors. Meanwhile, the applicability of this approach is unobvious in microcavities, which are multimode structures with a defect layer thickness of several microns, since in this case, a series of modes for any polarization inevitably arise within the band gap. Here, the Bragg frequency does not necessarily coincide with the frequency of any of the modes; in particular, it can be located in the free spectral range (wavelength separation between two successive reflected or transmitted optical intensity maxima). At present, multilayer photonic structures with the metal film are investigated within the concept of a Tamm plasmon-polariton (TPP) \cite{14,15,16}. \ins{ Its hybridization with microcavity modes are still an exotic object of study \cite{Kal,Bruk,Pankin}. Moreover, to the best of our knowledge, investigations of hybrid states of broadband TPP-microcavity modes have not been studied.} This approach could provide  new insight into the optical properties of the reflection-and-transmission PSs and expands the range of their application, in particular, for observing new physical effects and phenomena. In \cite{14, 15}, based on the coupled mode theory, the possibility  of localized state formation at the interface between a Bragg mirror and a thin metallic layer was theoretically predicted and experimentally demonstrated. As was shown in \cite{16}, the use of chromium as a metal leads to the excitation of a broadband TPP.   

In view of the aforesaid, in this study, we investigate the possibility of synchronous tuning of optical modes in the transmittance and reflectance spectra of a multimode reflection-and-transmission PS consisting of an asymmetric dielectric Fabry–P\'{e}rot microcavity and an ultrathin chromium film with the optical constants $n\approx k$. A dual-frequency nematic mixture is used as an electrically controlled structural element. The spectra of eigenmodes corresponding to the extraordinary waves are detected in the reflection and transmission simultaneously. The experimental data are compared with the results \ins{of the numerical simulation by Berreman $4\times4$-matrix technique}. 

\section{Experimental}

\begin{figure}
\centerline{\includegraphics[width=6cm]{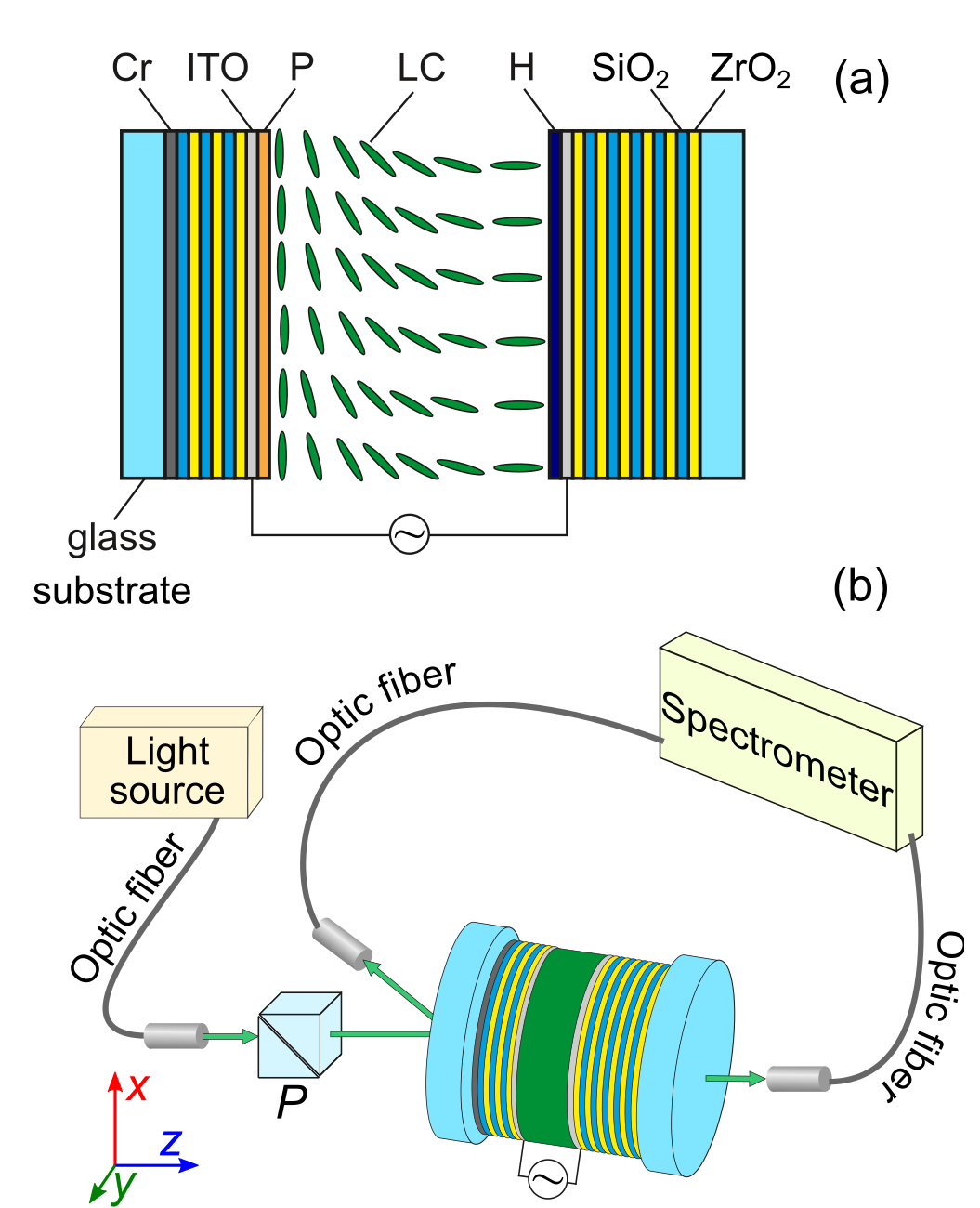}}
\caption{
(a) Configuration of the Cr-PS/LC cell. The LC hybrid configuration is implemented in the initial state ($U = 0$~V). Nematic mixture molecules are shown by elongated ellipsoids. (b) Scheme of an experimental setup for simultaneous recording of the transmittance and reflectance spectra of the photonic structure. Applied electric field \textbf{E} is directed along the $z$-axis, P is the polarizer. 
}
\label{fig1}
\end{figure}

The investigated PS consisted of two Bragg mirrors with an ultrathin metallic chromium (Cr) film ($\sim12$~nm) on the input mirror and a nematic LC layer between them as a structural defect: Sub/Cr(LH)$^m$--D--(HL)$^n$H (Fig.~\ref{fig1}a), where L (SiO$_2$) and H (ZrO$_2$) are the different dielectric optically isotropic layers with a low ($n_{\text{L}} = 1.45$) and high ($n_{\text{H}} = 2.05$) refractive indices, $m = 3$ and $n = 5$ are the numbers of the LH and HL bilayers (periods) to the left and right of defect layer D, respectively. Quartz glass substrates were used. A multilayer structure of mirrors was formed by alternate vacuum deposition of the ZrO$_2$ and SiO$_2$ oxides onto a substrate. According to the TEM data, the thicknesses of each layer were $d_{\text{L}} = (89 \pm 5)$~nm and $d_{\text{H}} = (66 \pm 5)$~nm. The periodicity of the structure produces a photonic band gap in the transmittance spectrum in the wavelength range of $420\div615$~nm. The presence of a metallic film on the input mirror forms a specific rejection band in the reflectance spectrum, the edges of which coincide with the PBG edges. Therefore, the overall reflectance profile $R(\lambda)$ of the PS acquires the form characteristic of the transmittance spectrum $T(\lambda)$. In turn, the periodicity violation leads to the appearance of resonance transmission and reflection peaks in both bands at the same wavelengths, which correspond to the optical modes localized on the defect layer. A dual-frequency nematic mixture MLC-2048 (Merck) was used as defect layer D. The dielectric anisotropy of this LC at $20^\circ$C is $\Delta\epsilon = +3.2$ at an applied electric field frequency of $f_1 = 1$~kHz and $\Delta\epsilon=-3.1$ at $f_2 = 50$~kHz \cite{12}. The nematic layer thickness was $d_{\text{LC}} = (7.0 \pm0.2)$~$\mu$m. In the initial state, a hybrid configuration of director \textbf{n} was formed in the nematic layer (Fig.~\ref{fig1}a). The hybrid alignment of director \textbf{n} is the distorted configuration created by the planar (P) orientation on the input mirror and the homeotropic (H) orientation on the other mirror. To obtain this configuration, the mirrors were coated with polyvinyl alcohol (PVA) (Sigma Aldrich) and surfactant N1,N2-Didodecyl-N1,N1,N2,N2-tetra-methylethane-1,2-diammoniumdibromide (Belarusian State Technological University), respectively. Transparent indium tin oxide (ITO) electrodes with a thickness of $\sim150$~nm deposited onto the surface of multilayers make it possible to control the structural transformations in the nematic by an electric field directed along the layer plane normal. 

The spectral positions of the transmission and reflection peaks of the Cr-PS/LC cell eigenmodes in the regimes of structural transformations “hybrid configuration -- quasi-homeotropic state” and “hybrid configuration -- quasi-planar state” in the nematic layer were experimentally investigated using setup schematically shown in Fig.~\ref{fig1}b. The thermostated sample was mounted such that the nematic director \textbf{n} on the input mirror was aligned along the $x$-axis of the laboratory system of coordinates ($x, y, z$). An ac voltage was applied to the Cr-PS/LC cell using a function generator AHP-3122 (AKTAKOM). Voltage $U$ applied to the sample was detected with a digit multimeter 34465A (Keysight Technologies). The polarized transmittance and reflectance spectra of the Cr-PS/LC cell were recorded using an Ocean Optics HR4000 spectrometer at a constant temperature of $t = (23.0\pm0.2)^\circ$C. A Glan prism used as a polarizer was installed in front of the sample such that electric field vector \textbf{E} of the incident light wave was parallel to the director \textbf{n} in the nematic surface layer at the input mirror. In this case, the resonance transmittance and reflectance peaks corresponding to the optical \textit{re}-modes are detected. These modes correspond to the traveling extraordinary (\textit{e}) waves with the refractive index changing along the propagation direction

\begin{equation} 
n_e(z)=\frac{n_\parallel n_\perp}
{\sqrt{n^2_\parallel\cos^2\theta(z)+n^2_\perp\sin^2\theta(z)}}
\label{eq1}
\end{equation}
                            
Here, $n_\parallel$ and $n_\perp$ are refractive indices of the LC for the incident radiation polarizations parallel ($\parallel$) and perpendicular ($\perp$) to the director \textbf{n} of a homogeneous nematic layer and $\theta(z)$ is the angle between the wave vector $\textbf{k} \parallel z$ and the local direction of the director \textbf{n}. Since, during the structural transformations in the nematic, the distribution of the angle $\theta(z)$ and the effective refractive index of the LC medium change, then the \textit{re}-modes of the Cr-PS/LC cell, in contrast to the \textit{ro}-modes ($n_o=n_\perp=\text{const}$), become sensitive to the electric field. Below, we only consider the behavior of the modes of this type. The radiation of a broadband light source was introduced into the sample at an angle of $\sim4^\circ$ and coupled out to the spectrometer using optical fibers equipped with collimators. A small angle of incidence of the probe radiation ensured the simultaneous recording of the transmittance and reflectance spectra of the PS.

\section{Results and Discussions}

\begin{figure}
\centerline{\includegraphics[width=6.4cm]{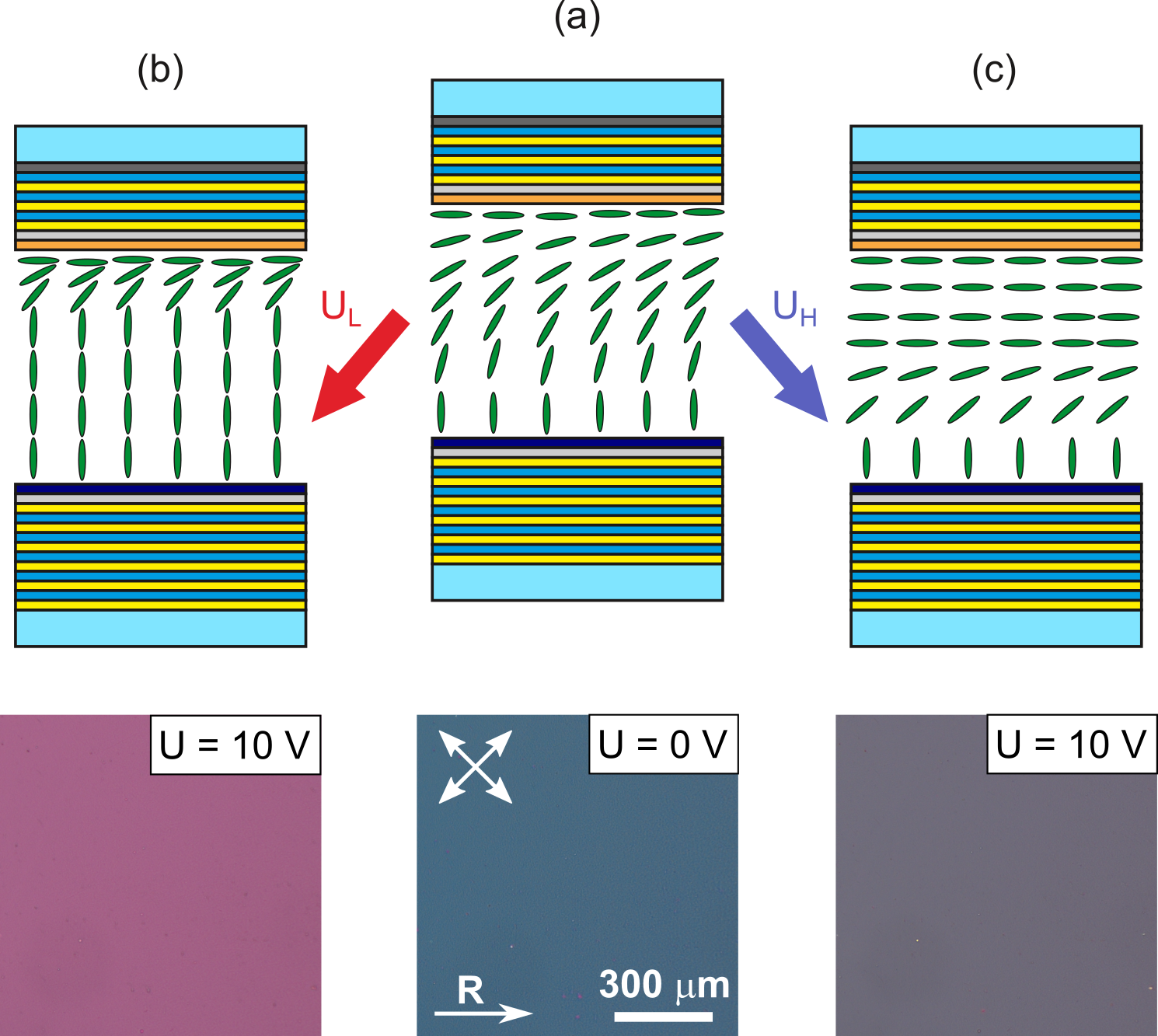}}
\caption{
Configurations of LC director \textbf{n} (top row) and microphotographs of optical textures of the Cr-PS/LC cell in the crossed polarizers geometry (bottom row): (a) hybrid configuration, (b) quasi-homeotropic state ($f = 1$~kHz), and (c) quasi-planar state ($f = 50$~kHz). {\bf R} is the rubbing direction of the PVA film. Double arrows show the polarizer and analyzer directions. 
}
\label{fig2}
\end{figure}

\begin{figure}
\begin{center}
\includegraphics[width=6.5cm]{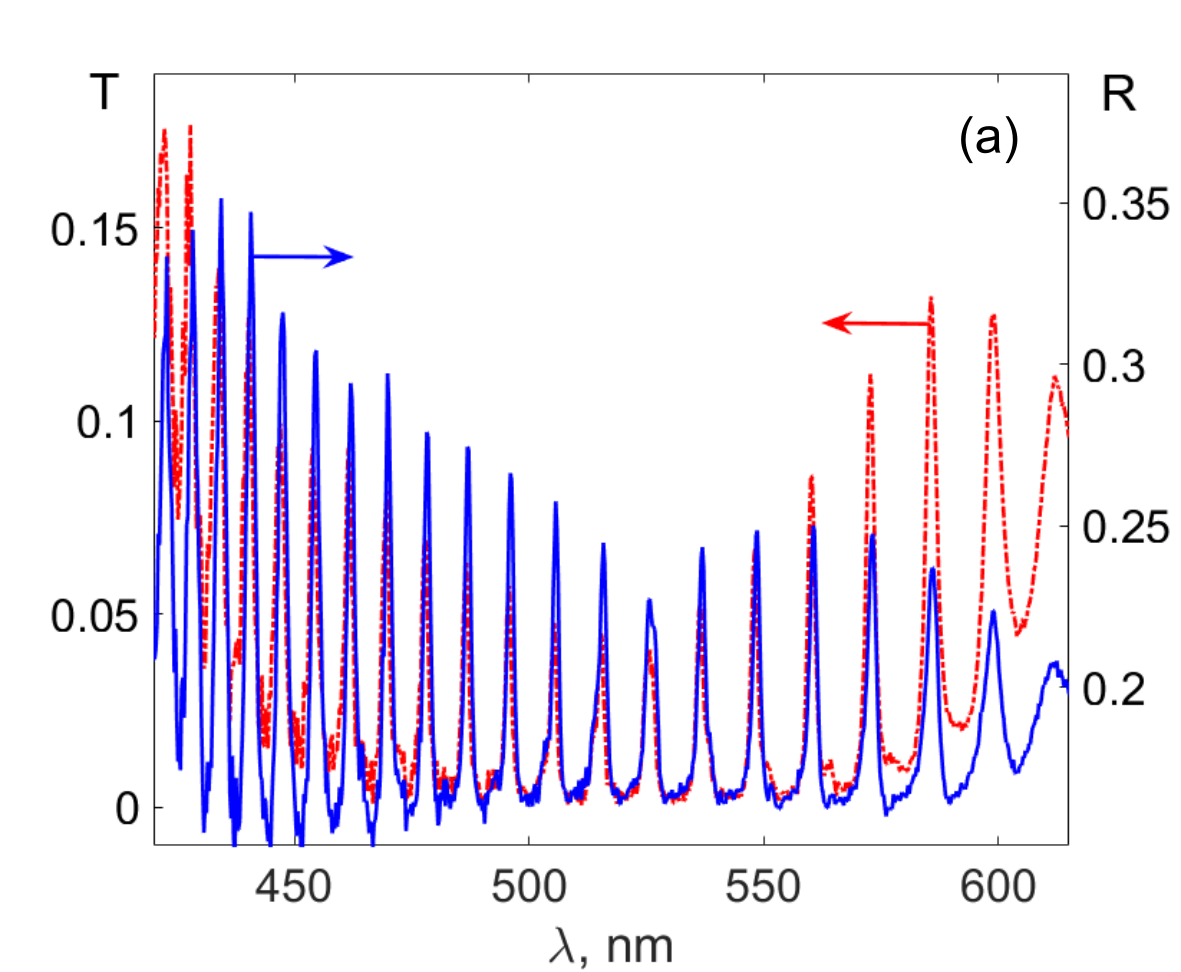}
\includegraphics[width=6.5cm]{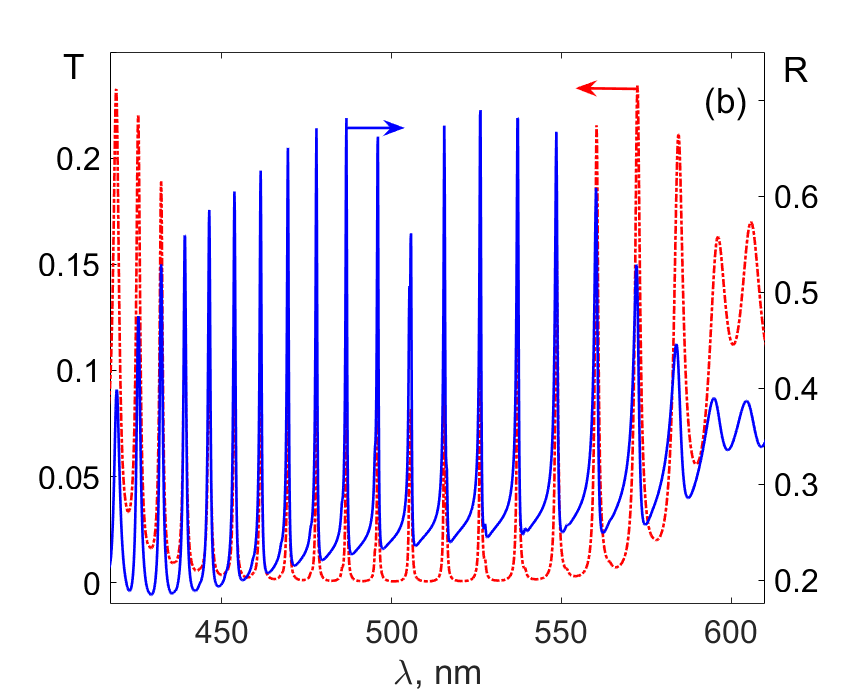}
\end{center}
\caption{
Transmittance spectra $T(\lambda)$ (red dash-and-dot lines) and reflectance spectra $R(\lambda)$ (blue solid lines) of the Cr-PS/LC cell corresponding to the re-modes at zero voltage ($U = 0$~V): (a) experimental data and (b) results of the numerical simulation. 
}
\label{fig3}
\end{figure}

An ac voltage of different frequencies in the range of ($0-10$)~V applied to the sample induces a thresholdless reorientation of the director \textbf{n} in the LC bulk from the initial hybrid configuration shown in Fig.~\ref{fig2}a (top row). In particular, the low-frequency ($f = 1$~kHz) voltage $U_\text{L}$, due to the positive dielectric anisotropy of the nematic, reorients the director \textbf{n} along the electric field orthogonally to the multilayers and, at the maximum voltage of $U = 10$~V, the director configuration has the form shown in Fig.~\ref{fig2}b (top row). The reorientation of the director perpendicular to the electric field occurs at the high-frequency ($f = 50$~kHz) voltage $U_\text{H}$, because, at this frequency, the nematic has the negative dielectric anisotropy (Fig.~\ref{fig2}c, top row). The orientation effects modify the optical response of the Cr-PS/LC cell, which can be seen in microphotographs of the PS optical textures in the crossed polarizer geometry (Fig.~\ref{fig2}, bottom row). The texture in Fig.~\ref{fig2}b at a voltage of $U_\text{L}$ = 10~V corresponds to the quasi-homeotropic state and the texture in Fig.~\ref{fig2}c at a voltage of $U_\text{H}$ = 10~V corresponds to the quasi-planar state. Thus, the color of each texture is determined by the director configuration with allowance for the general feature of light transmission through the PS with the PBG. The homogeneity of the textures shown in Fig.~\ref{fig2} and the textures obtained at the intermediate voltages shows that the director alignment uniformly changes with increasing electric field in one plane perpendicular to the mirrors over the entire bulk of the LC layer. Figure~\ref{fig3} shows the experimental (Fig.~\ref{fig3}a) and calculated (Fig.~\ref{fig3}b) polarized transmittance $T(\lambda)$ and reflectance $R(\lambda)$ spectra of the Cr-PS/LC cell at $U = 0$~V, when the orientation of the MLC-2048 nematic mixture corresponds to the hybrid configuration of the director (Fig.~\ref{fig1}a). It can be seen that the spectra of the investigated structure are sets of resonance peaks within the coinciding rejection band and PBG. The $T$- and $R$-spectra simultaneously recorded in the experiment (Fig.~\ref{fig3}a) shows that each transmission peak coincides with the corresponding reflection peak. This means that both peaks correspond to the same $re$-mode, which resonates in the defect layer at a certain frequency.
 
The numerical simulation of the spectra of the investigated structure (Fig.~\ref{fig3}b) was carried out by the \ins{Berreman $4\times4$-matrix technique} \cite{17}.  The simulation was carried out taking into account the dispersion properties of the Cr-PS/LC cell materials \cite{18,19,20,21,22} and with the parameters of the nematic mixture MLC-2048. In particular, splay and bend elastic constants of K$_{11} = (15\pm1)$~pN and K$_{33}=(20\pm1)$~pN obtained by the Frederiks transition measurements on pure MLC-2048 at $25^\circ$C \cite{23} were used. In this case, the distribution of the director \textbf{n} over the thickness of the hybrid cell with the nonsymmetric surface angles at a director angle $\theta$ from $0^\circ$ to $90^\circ$ and a ratio of K$_{33}/\text{K}_{11}\sim1.3$ between elastic constants of the nematic is almost linear \cite{24}. Therefore, the variable refractive index $n_\perp \leqslant n_e(z)\leqslant n_\parallel $ (Eq.~\ref{eq1}) was calculated using the linear angular distribution $\theta(z)$ between wave vector \textbf{k} and the local director \textbf{n}. It can be seen in Fig.~\ref{fig3} that the experimental and calculated spectral positions of the modes of the investigated structure are in good agreement with each other. 
\ins{It turned out that the observed discrepancy between the distributions of amplitudes of the experimental and calculated reflection peaks cannot be  eliminated by taking into account  losses caused by  imperfect multilayer structure of the mirrors, presence of ITO and alignment layers. Probably, the presence of hybrid states of broadband TPP and microcavity modes leads to the appearance of some loss factors which are difficult to reveal.} 

The Cr-PS/LC cell is, in fact, an asymmetric microcavity with an ultrathin metallic film on the input mirror; therefore, the resonance condition known from the Fabry–P\'{e}rot theory 
\begin{equation} 
\lambda_e=\frac{2\langle n_e\rangle d}
{m_e}
\label{eq2}
\end{equation}
\begin{figure}
\begin{center}
\includegraphics[width=6.5cm]{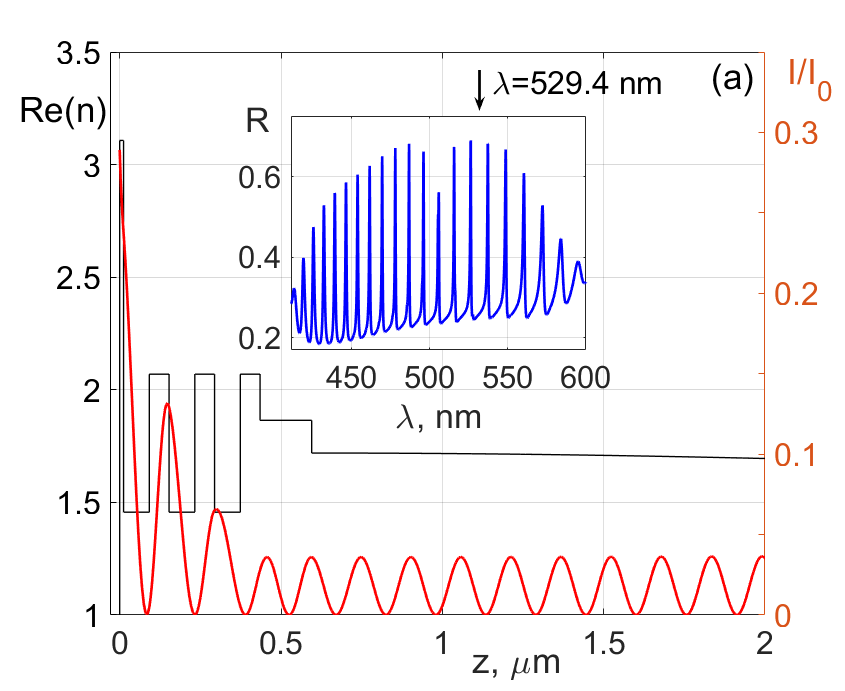}
\includegraphics[width=6.5cm]{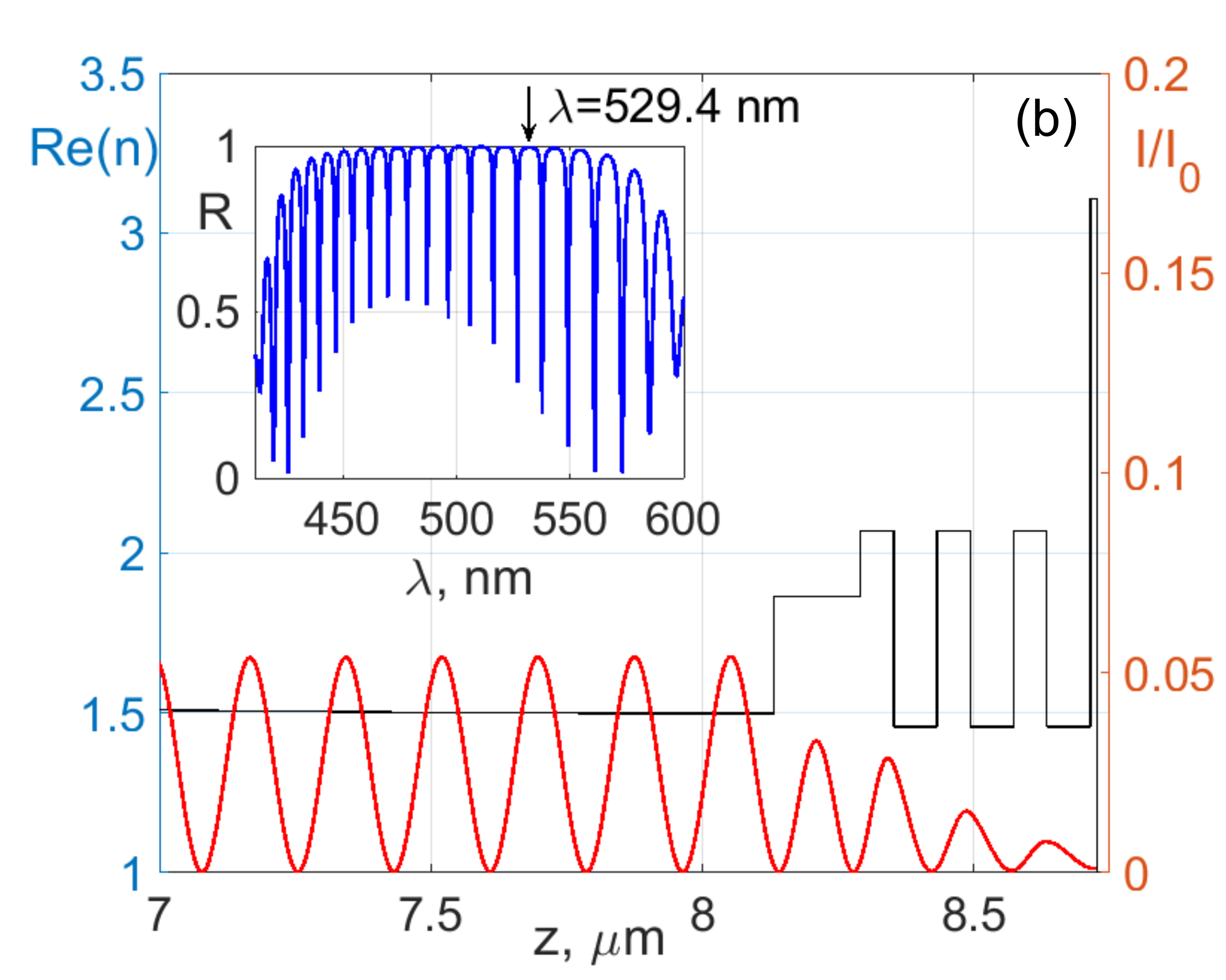}
\end{center}
\caption{
Distribution of the light field intensity at a plasmon-polariton wavelength of 529.4~nm (shown by the arrow) normalized to the input intensity (red lines) and spatial distribution of the refractive index of the structure layers (black lines) under illumination of the Cr-PS/LC cell from (a) the metallic film side and (b) from the Bragg mirror side. Insets: corresponding calculated reflectance profiles $R(\lambda)$. 
}
\label{fig4}
\end{figure}
is applicable to the investigated structure. Here, $\langle n_e\rangle=(1/d)\int_0^d n_e (z)dz$ is the effective refractive index of the LC medium (angle brackets indicate averaging over the layer thickness); integers me denote the numbers of defect modes corresponding to the number of standing wave antinodes in the cavity. For the sake of simplicity, the phase change upon reflection from the mirrors is ignored here \cite{25}. Thus, the profile of the transmittance spectrum of the Cr-PS/LC cell almost does not differ from that for the all-dielectric Fabry––P\'{e}rot cavity. On the contrary, the reflectance spectrum of the Cr-PS/LC cell has the features: instead of a wide reflection band, a rejection band is formed, in which narrow reflection peaks appear at the resonant frequencies.  \ins{The rejection band} originates from the  \ins{formation} of a broadband TPP at the metal/Bragg mirror interface when the structure is illuminated from the metal side \cite{16}. 
\ins{Figure~\ref{fig4}a shows  the light field intensity distribution at a plasmon-polariton wavelength of 529.4~nm  under illumination of the Cr-PS/LC cell from  the metallic film side.}
The plasmon-polariton wavelength  is determined from the phase matching condition  $|r_m|e^{i\phi_{m}}|r_{bm}|e^{i\phi_{bm}}=1$ \cite{14}, where $\phi_m$, $r_m$, and $\phi_{bm}$, $r_{bm}$ are the phases and amplitude coefficients of the waves reflected from the metallic film and the Bragg mirror, respectively. \ins{As can be seen in Fig.~\ref{fig4}a, TPP wavelength is located}  within the free spectral range in the vicinity of the rejection band center. The calculated distribution of the light field intensity at this wavelength shown in Fig.~\ref{fig4}a   \ins{demonstrates} the field localization at the metal/Bragg mirror interface ($z\approx0~\mu$m) \ins{that evidences for  TPP excitation.}
\begin{figure}
\begin{center}
\includegraphics[width=6.5cm]{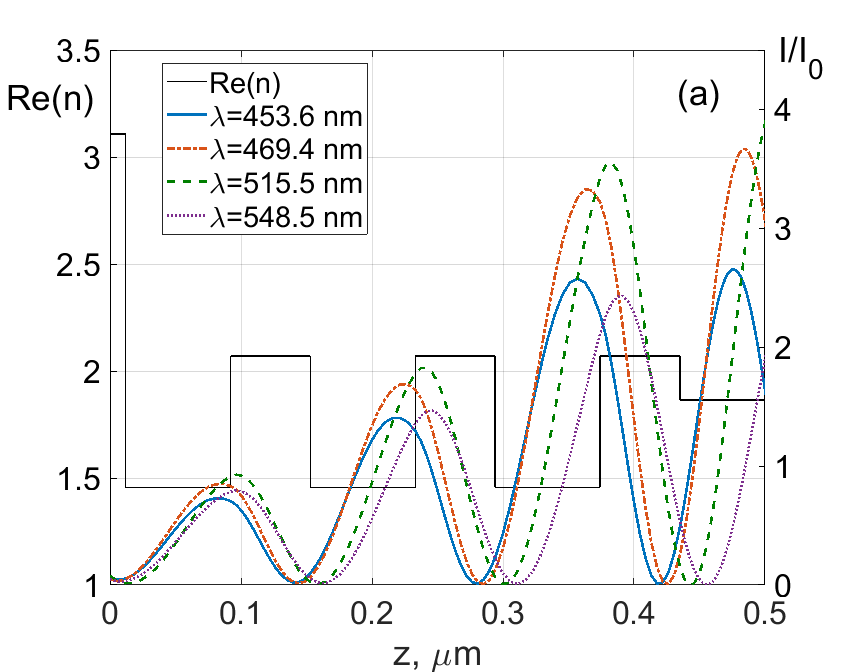}
\includegraphics[width=6.5cm]{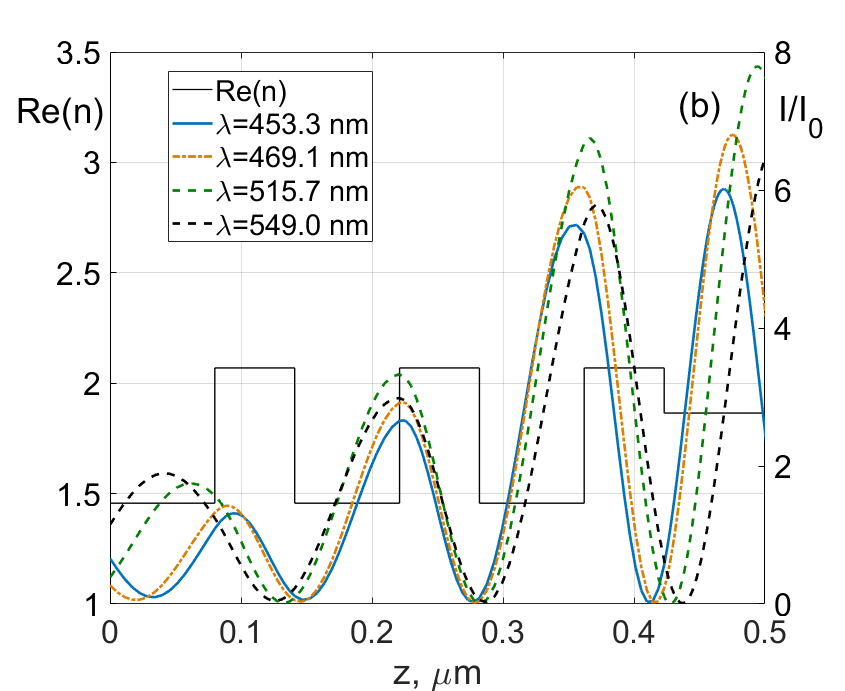}
\end{center}
\caption{Distribution of the light field intensity for arbitrarily selected modes of photonic structure with (a) and without (b) chromium film, and spatial distribution of the refractive index of the structure layers (black lines).}
\label{fig5}
\end{figure}
The strong absorption induced by the metal provides a fairly high rejection level in reflection of the off-resonant radiation in a wide spectral range, which almost coincides with the PBG. \ins{For comparison, Fig.~\ref{fig4}b shows the light field intensity distribution (at the same wavelength $\lambda =529.4$~nm) calculated under illumination of the structure from the Bragg mirror side. In this case, despite the presence of a metallic film, the TPP is not excited and the reflectance spectrum of the structure becomes complementary to the transmittance spectrum: instead of peaks, there are dips against the background of a wide reflection band.}

On the other hand, the appearance of reflection peaks in the rejection band at the resonance frequencies is  \ins{caused by hybridization of TPP and microcavity modes which leads to the specific spatial distribution of wave fields (Fig.~\ref{fig5}).} 
\ins{ For such distribution the nodes of standing waves of presented modes are localized on the metal layer (Fig.~\ref{fig5}a).
In this case, the absorption factor of the reflected radiation is negligible \cite{6}. Hybridization of TPP and microcavity modes lead to a drastic change in the reflection spectrum of the photonic structure containing metal layer. In particular, the reflection peaks  are observed instead of dips in the rejection band at the resonance frequencies. Obviously, this situation will be realized for any mode in the band gap. Such a specific wave field distribution at the resonance frequencies is apparently retained even upon variation in the refractive index of the LC medium, which makes it possible, as we show below, to implement the synchronous tuning of the reflection and transmission peaks during the structural transformations in the nematic.
In the absence of the metal film,  localization of the standing wave nodes at $z\approx0~\mu$m disappears (Fig.~\ref{fig5}b). }  

\begin{figure}
\begin{center}
\includegraphics[width=4.2cm]{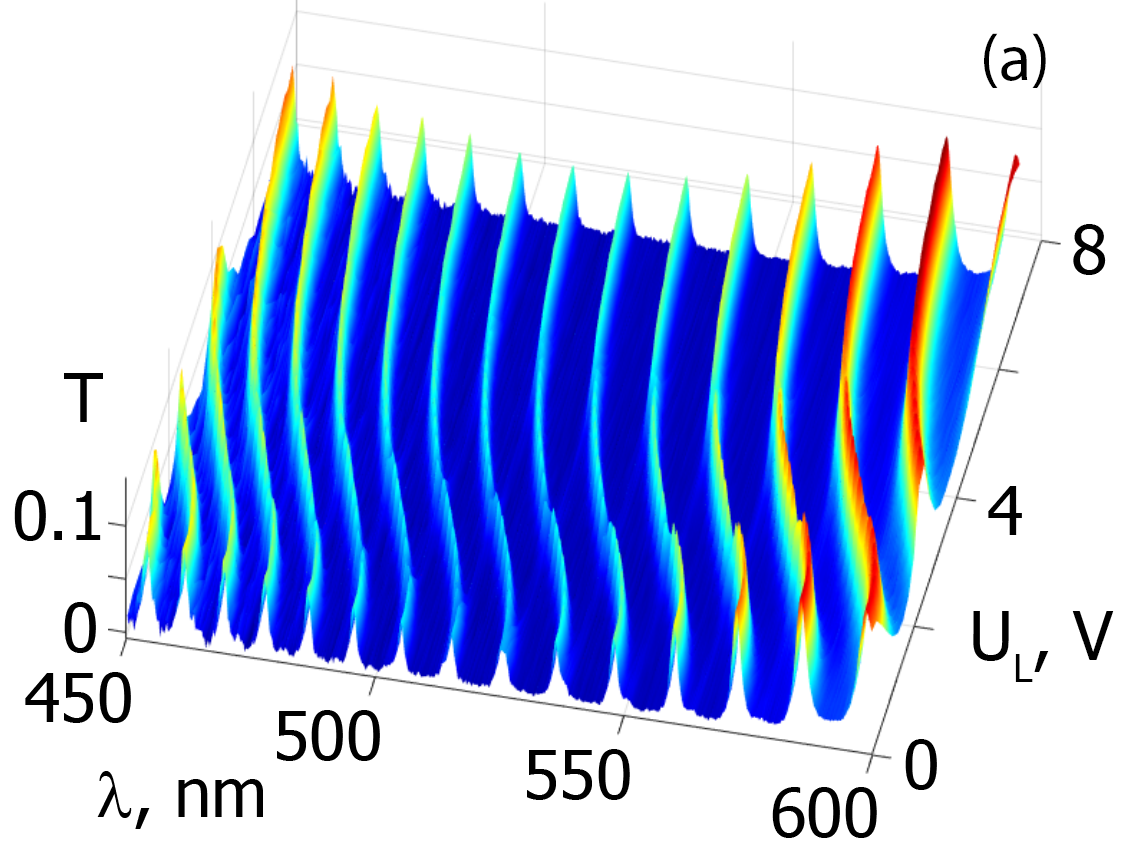}
\includegraphics[width=4.2cm]{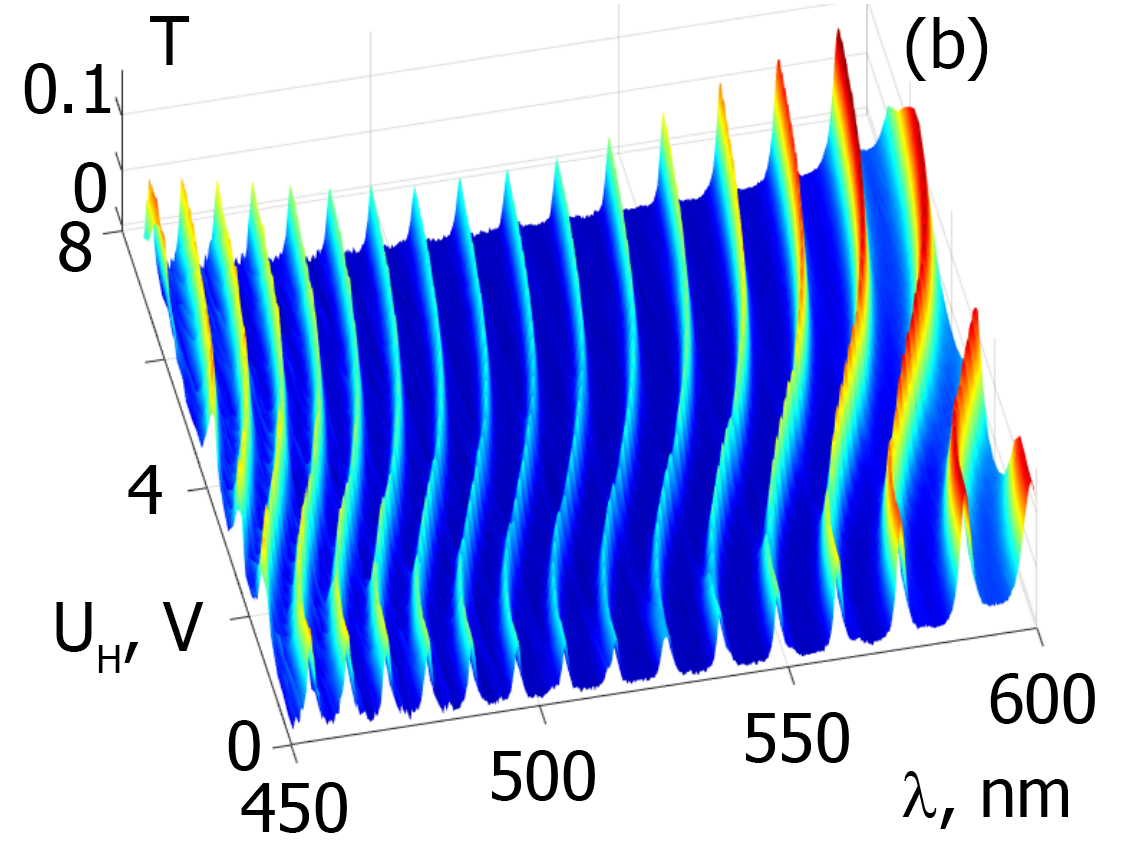}
\includegraphics[width=4.2cm]{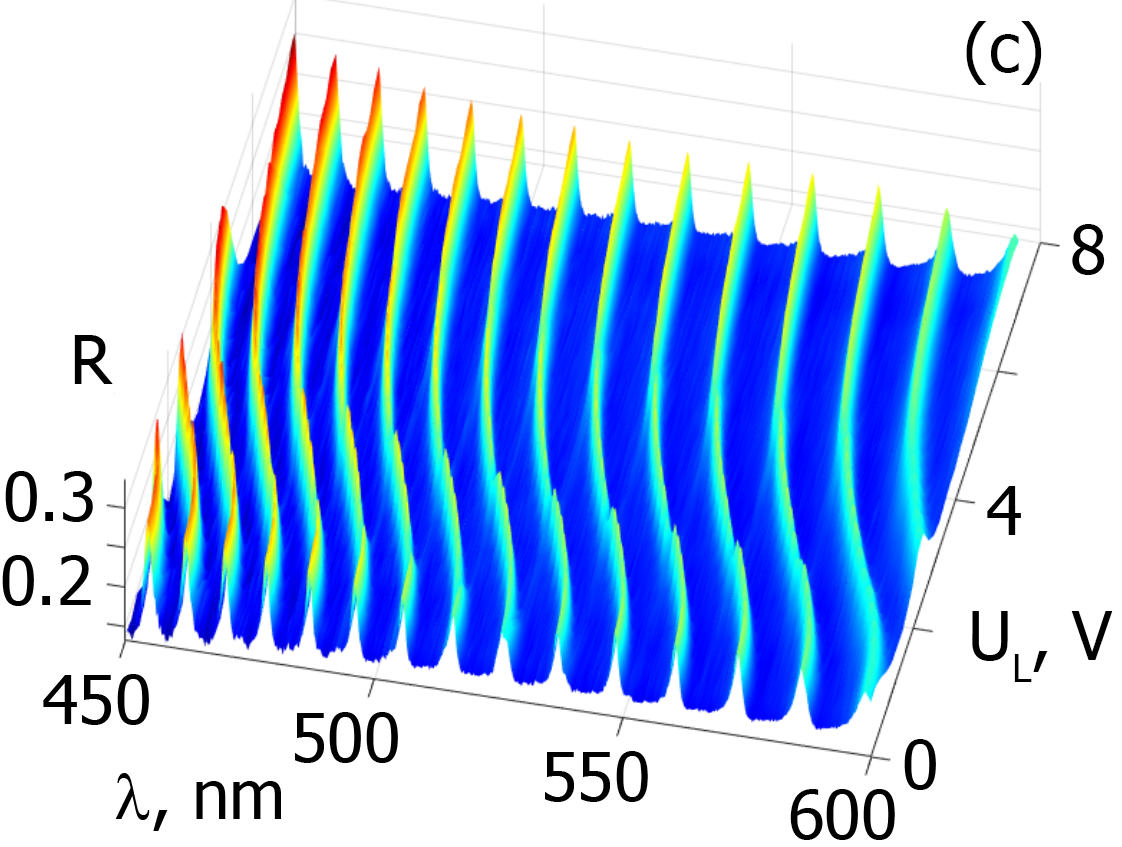}
\includegraphics[width=4.2cm]{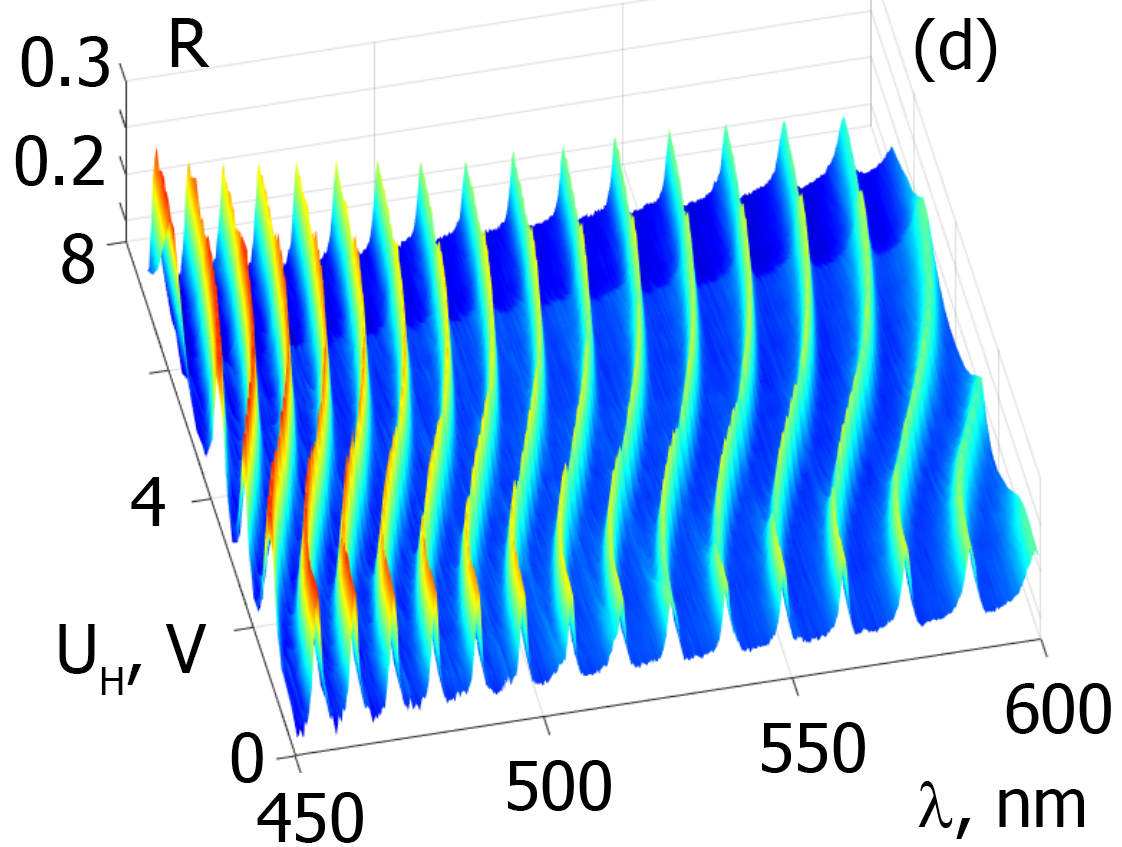}
\end{center}
\caption{
3D-patterns of the Cr-PS/LC cell consisting of a set of (a, b) transmittance spectra $T(\lambda)$ and (c, d) reflectance spectra $R(\lambda)$. $T$- and $R$-spectra were recorded simultaneously at (a, c) the low-frequency voltage $U_\text{L}$ or (b, d) the high-frequency voltage $U_\text{H}$ in the range of ($0 \div 7$)~V with a step of 0.25~V.  
}
\label{fig6}
\end{figure}

Figure~\ref{fig6} shows 3D patterns of the transformation of the experimental transmittance $T(\lambda)$ and reflectance $R(\lambda)$ spectra of the Cr-PS/LC cell with a dual-frequency nematic mixture under the low-frequency voltage $U_\text{L}$ (Fig.~\ref{fig6}a, c) and the high-frequency voltage $U_\text{H}$ (Fig.~\ref{fig6}b, d). Each pair of the $T$- and $R$-spectra was recorded simultaneously, first at the low-frequency and then at the high-frequency voltage in the range of ($0 \div 7$)~V with a step of 0.25~V. Depending on the frequency of the applied voltage, a smooth reorientation of the director \textbf{n} by an angle of up to $90^\circ$ ($U = U_\text{L}$) or by an angle of up to $0^\circ$ ($U = U_\text{H}$) in the ($xz$) plane occurs in the defect LC layer. In the first case, the effective refractive index of the LC medium decreases $\langle n_e\rangle \to n_\perp$ and, in the second case, increases $\langle n_e\rangle \to n_\parallel$. Then, according to resonance condition Eq.~\ref{eq2}, the transmission and reflection peaks will shift to the short-wavelength (at $U_\text{L}$) or the long-wavelength (at $U_\text{H}$) spectral range with increasing voltage. Since each pair of the $T$- and $R$-peaks coinciding in wavelength corresponds to one mode with certain number $m_e$, the peaks will shift synchronously. It can be seen in Fig.~\ref{fig6} that the transformation of the $T$- and $R$-spectra fully shows the above-mentioned features of both the resonance properties of the entire PS and the electrically induced structural transformations in the defect LC layer. When the voltage is switched off, the nematic layer returns to its initial state corresponding to the hybrid configuration. At that, the $T$- and $R$-modes occupy their initial spectral positions.

\begin{figure}
\centerline{\includegraphics[width=6cm]{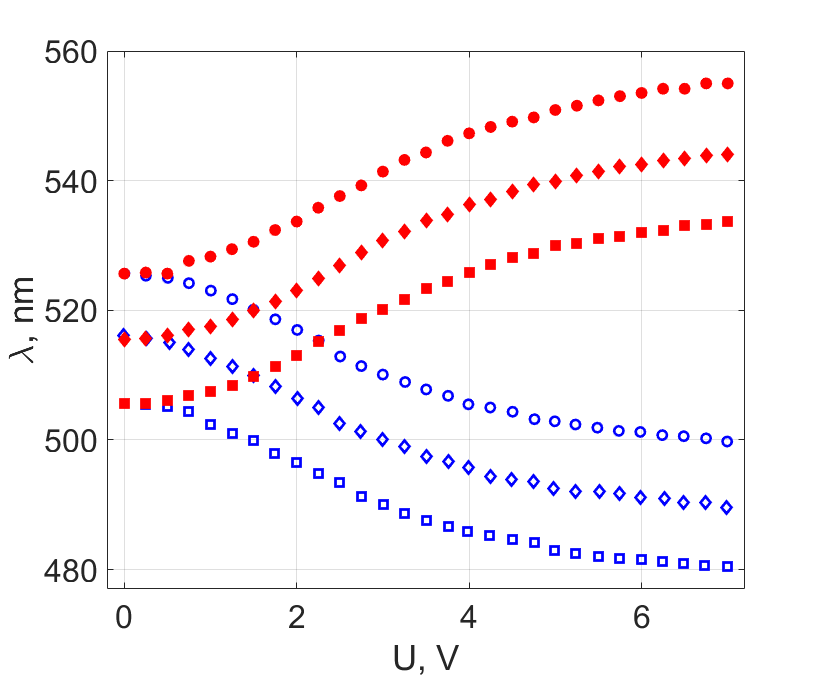}}
\caption{
Spectral positions of the maxima of the coinciding $T$- and $R$-peaks for three $re$-modes at wavelengths of 505.7~nm (open and closed squares), 515.3~nm (open and closed diamonds), and 525.5~nm (open and closed circles) as a function of the applied low-frequency (blue symbols) and high-frequency (red symbols) voltage.
}
\label{fig7}
\end{figure}

Figure~\ref{fig7} shows field dependencies of the spectral positions of the maxima of three re-modes of the Cr-PS/LC cell at the PBG center in the voltage range of ($0 \div 7$)~V. Here, each point corresponds to two experimental wavelengths for the transmission and reflection peaks (Fig.~\ref{fig6}). Good agreement between these values at each step of the applied voltage is indicative of synchronous tuning of the $T$- and $R$-peaks. The form of these dependencies reflects the thresholdless character of the structural transformations “hybrid configuration $\to$ quasi-homeotropic state” and “hybrid configuration $\to$ quasi-planar state” in the nematic defect layer of the PS. In the considered voltage range, each pair of peaks shifts from its initial position to the red (upper curves) or blue (lower curves) spectral regions, depending on frequency. The shift value varies from zero to $\sim 26$~nm, which corresponds to approximately two and a half free spectral range. It can be seen that the $\lambda(U)$ curves symmetrically diverge with increasing voltage. In this case, the interval between positions of the same mode at a fixed low- or high-frequency voltage is doubled. For example, the central mode 515.3~nm ($U = 0$) occupies the 490~nm position at the low frequency ($U_\text{L} = 7$~V) or the 542~nm position at the high frequency ($U_\text{H} = 7$~V). Thus, the monotonic behavior of the $\lambda(U)$ dependencies allows one to smoothly adjust the range of synchronous switching of the transmission and reflection peaks corresponding to any mode by changing a value of voltage $U$ applied to the sample and to implement the switching itself by changing the operating frequency (1~kHz~$\leftrightarrow$~50~kHz).

\section{Conclusion}

The electro-optical properties of the Fabry–P\'{e}rot cavity-type multilayer photonic structure based on the distributed Bragg mirrors with an ultrathin chromium film were studied. The dual-frequency nematic mixture MLC-2048 with the hybrid configuration of the director in the initial state was used as a defect in the ZrO$_2$/SiO$_2$ periodic structure. In contrast to the transmission, the reflectivity of the structure substantially depends on the incident radiation direction. Under illumination from the side of the second Bragg mirror, the profile of the reflection spectrum $R(\lambda)$ is typical of the all-dielectric Fabry–P\'{e}rot cavity. In particular, a broad reflection band with narrow dips at the defect mode frequencies within the PBG is observed. Under illumination from the metallic film side, the reflectance spectrum $R(\lambda)$ is reversed. Specifically, instead of the reflection band, a rejection band with the resonance $R$-peaks appears. Thus, the profile of the reflectance spectrum of the structure becomes similar to the profile of the $T(\lambda)$ spectrum. In this case, the spectral positions of the transmission and reflection peaks corresponding to the same mode coincide. The calculation of the wave field distribution for both cases showed that the reversal of the reflection spectrum in the second case is caused by the excitation of a broadband Tamm plasmon-polariton at the metal/Bragg mirror interface. Based on the electric field-induced structural transformations “hybrid~$\to$~quasi-homeotropic state” and “hybrid~$\to$~quasi-planar state” in the nematic defect layer, a synchronous tuning of the transmission and reflection peaks of the photonic structure was implemented. The ratio between the amplitudes of the coinciding $T$- and $R$-peaks is tuned by changing the number of periods in the (LH)$^m$/(HL)$^n$ structure \cite{6}. Depending on the frequency of the voltage applied to the sample, the peaks shift to both the blue and red spectral regions. The monotonicity of the $\lambda(U)$ dependencies reflects the process of structural transformations in the nematic layer. The shift value is controlled by the value of the applied voltage. At a maximum operating voltage of 7~V, the shift of the peaks in both directions for used defect layer thickness is about two and a half free spectral range. In addition, switching the frequency 1~kHz~$\leftrightarrow$~50~kHz of any fixed voltage makes it possible to implement synchronous switching of the spectral position of the photonic structure modes in the transmittance and reflectance spectra. In this case, the width of the interval for switching modes from one extreme position to another depends on the value of the applied voltage.

\end{document}